\documentstyle[preprint,aps,graphicx]{revtex}
\tightenlines

\newcommand{\r} {{\mathbf r}}
\newcommand{\si} {\sigma {\mathrm i}}
\newcommand{\ns} {n_\sigma}

\newcommand{\fsi} {f_{\sigma {\mathrm i}}}
\newcommand{\psisi} {\psi_{\sigma {\mathrm i}}}
\newcommand{\Vkseffs} {V^{\mathrm{KS}}_{{\mathrm{eff},\sigma}}}
\newcommand{\Vext} {V_{\mathrm{ext}}}
\newcommand{\Vh} {V_{\mathrm{H}}}
\newcommand{\Vxcs} {V_{\mathrm{xc},\sigma}}
\newcommand{\Ns} {N_{\sigma}}
\newcommand{\oo} {object-oriented}
\newcommand{\Oo} {Object-oriented}
\newcommand{\F} {Fortran}
\newcommand{\simbox} {{\bf simbox}}
\newcommand{\grid} {{\bf grid}}
\newcommand{\subgrid} {{\bf subgrid}}
\newcommand{\new} {{\bf new}}
\newcommand{\delete} {{\bf delete}}
\newcommand{\display} {{\bf display}}
\newcommand{\simboxconstruct} {{\bf simbox$\_$construct}}
\newcommand{\simboxprint} {{\bf simbox$\_$print}}
\newcommand{\gridconstruct} {{\bf grid$\_$construct}}
\newcommand{\griddestruct} {{\bf grid$\_$destruct}}
\newcommand{\gridprint} {{\bf grid$\_$print}}
\newcommand{\subgridconstruct} {{\bf subgrid$\_$construct}}
\newcommand{\subgriddestruct} {{\bf subgrid$\_$detruct}}
\newcommand{\wavefunction} {{\bf wavefunction}}
\newcommand{\Msimbox} {{\bf m$\_$simbox}}
\newcommand{\Mgrid} {{\bf m$\_$grid}}
\newcommand{\Msubgrid} {{\bf m$\_$subgrid}}
\newcommand{\MODULE} {{\bf MODULE}}
\newcommand{\TYPE} {{\bf TYPE}}
\newcommand{\USE} {{\bf USE}}
\newcommand{\ONLY} {{\bf ONLY}}
\newcommand{\PUBLIC} {{\bf PUBLIC}}
\newcommand{\PRIVATE} {{\bf PRIVATE}}
\newcommand{\CONTAIN} {{\bf CONTAIN}}
\newcommand{\MODULEPROCEDURE} {{\bf MODULE PROCEDURE}}
\newcommand{\FUNCTION} {{\bf FUNCTION}}
\newcommand{\SUBROUTINE} {{\bf SUBROUTINE}}
\newcommand{\POINTER} {{\bf POINTER}}
\newcommand{\TARGET} {{\bf TARGET}}
\newcommand{\INTENT} {{\bf INTENT}}
\newcommand{\IN} {{\bf IN}}
\newcommand{\ALLOCATABLE} {{\bf ALLOCATABLE}}
\newcommand{\ALLOCATE} {{\bf ALLOCATE}}
\newcommand{\DEALLOCATE} {{\bf DEALLOCATE}}
\newcommand{\NULLIFY} {{\bf NULLIFY}}
\newcommand{\mg} {multigrid}

\begin{document}
\draft

\title{Object-oriented construction of a multigrid electronic-structure code
	   with Fortran 90}

\author{Yong-Hoon Kim,$^1$\footnote{
	{\it Corresponding author}\\
	Address: Department of Physics, 1110 W. Green St., Urbana, IL 61801-3080, USA\\ 
	E-mail: yhoon@physics.uiuc.edu\\
	Telephone number: (217) 244-0391\\
	Fax number: (217) 333-9819}
	In-Ho Lee,$^2$ 
	and Richard M. Martin$^{1,3}$}

\address{$^1$ Department of Physics,
	University of Illinois at Urbana-Champaign,
	Urbana, IL 61801, USA}
\address{$^2$ School of Physics, Korea Institute for Advanced Study,
	Cheongryangri-dong, Dongdaemun-gu, Seoul 130-012, Korea}
\address{$^3$ Material Research Laboratory,
	University of Illinois at Urbana-Champaign,
	Urbana, IL 61801, USA}

\date{October 7, 1999}

\maketitle

\begin{abstract}
We describe the \oo\ implementation of a higher-order
finite-difference density-functional code in \F\ 90.  \Oo\ models of
grid and related objects are constructed and employed for the
implementation of an efficient one-way \mg\ method we have recently
proposed for the density-functional electronic-structure calculations.
Detailed analysis of performance and strategy of the one-way \mg\
scheme will be presented.

\vspace*{0.3cm}


Object-oriented programming; Fortran 90; Scientific computing
\end{abstract}

\narrowtext
\section{Introduction}
In recent years,  the usefulness of the real-space technique based on
three-dimensional uniform grid and accurate forms of finite-difference
formula has been demonstrated for the electronic-structure calculations
\cite{Che94,Bri95,IHL,YHK,Anc99,Wan99}. Since all the computations are
performed in real space, the major part of the calculations are local
operations, which makes the algorithm easily parallelized and
implementation of order-N algorithm \cite{Goe99} transparent. This is
in contrast to plane wave method methods which rely upon global fast
Fourier transforms. Furthermore, since the Laplacians and
potential-wave functions multiplications are respectively evaluated by
the finite-difference operation on the wave functions and a simple
one-dimensional vector multiplications on the fly, explicit storage of
the Hamiltonian matrix elements can be avoided and the matrix
diagonalization can be efficiently performed by iterative
diagonalization methods such as conjugate gradient method.

Although there has been great emphasis on the algorithmic developments
in the real-space electronic-structure calculation schemes, we feel
that the issue of the code construction and organization has been
relatively neglected.  Recently, we have proposed an efficient and
easy-to-implement one-way \mg\ algorithm which results in significant
enhancement in computation speed for grid-based iterative
electronic-structure calculations \cite{IHL3}.  In spite of the
advantage of the \mg\ in general \cite{Hea97,Pre92,Dou96}, it requires
complications of coding and organization of data structure because it
involves several grid levels and data transfer between them.  In such
a situation, we found the modern programming construction paradigm,
\oo\footnote{Although ``object-orientation'' is both a language feature
and a design methodology, this paper is primarily concerned with the
design aspect.} programming \cite{Rum91} can be useful.  Our original
code was written in \F\ 90 \cite{Ell94} with heavy recycling of our
previous planewave codes written in \F\ 77, in conventional non-\oo\
programming style, with the purpose of typical scientific programming,
namely the quick implementation of a physical idea.  Although we tried
to keep the code maintainable and clear, it quickly became long and
messy with each addition of function and implementation of an idea.
We recognized that the complexity of the conventional programming
style is an obstacle, or at least a complication, especially for the
implementation of \mg .  Hence, we restructured the code by
introducing \oo\ modeling concepts, and in this work we will report
our experience of this transition and the implementation of the \mg\
method with the newly designed code.  Although \oo\ programming should
be most straightforward in \oo\ languages such as C++, it is in
principle also possible in non-\oo\ languages \cite{Rum91}, and
especially relatively easy with \F\ 90 which supports many ingredients
of \oo\ coding.  \Oo\ scientific programming using \F\ 90 has been of
much interest in recent years \cite{Dec97,Car97,Gra97,Dec98,Dub99},
and this paper will add additional information to this discussion.  In
addition, since grid-based simulations are common in other scientific
and engineering computations, we expect our work is a helpful guide for
the code construction in those fields.

The outline of the current paper is as follows.  In
Sec. \ref{sec:KS-DFT}, we first describe the electronic-structure
calculation scheme within the Kohn-Sham density-functional theory, and
our methodology based on the higher-order finite difference
formulation.  In Sec. \ref{sec:F90-OOP}, we review the key concepts of
\oo\ methodologies, and describe the introduction of \oo\ concepts
into our grid-based program written in \F\ 90.  In
Sec. \ref{sec:OOP-MG}, we briefly review the \mg\ theory, and describe
the one-way \mg\ scheme which we have recently
proposed\cite{IHL3}. Implementation of the one-way \mg\ method is
discussed, and especially the simplification induced by the \oo\ design
is emphasized. The impressive enhancement of computational efficiency
due to the introduction of the one-way \mg\ method has been
demonstrated in our previous publication, and here more detailed
analysis of performance test and \mg\ strategy will be reported.  The
current work will be summarized in Sec. \ref{sec:conclusions}.

\section{Higher-order finite-difference Kohn-Sham
electronic-structure calculation method}
\label{sec:KS-DFT}

The (spin-dependent) Kohn-Sham (KS) density-functional theory (DFT)
\cite{KS-DFT} is an independent-electron theory in which one obtains
the single-particle wave functions $\psi_{\si}(\r)$ for spin channel
$\sigma = \uparrow, \downarrow$ and eigenvalues $\epsilon_{\si}$ by
solving KS equations (Hartree atomic units are used throughout the
paper)
\begin{equation}
 \biggl[ -\frac{1}{2} \nabla^2 + \Vkseffs(\r) \biggr] \psisi (\r) =
 \epsilon_{\si} \psi_{\si}(\r),
\label{eq:KS-eq}
\end{equation}
with the spin density
\begin{equation}
 \ns(\r) = \sum_{i=1}^{\Ns} \fsi |\psisi(\r)|^2,
\label{eq:density}
\end{equation}
where $\Ns$ is the number of occupied $\sigma$ spin orbitals. The effective
KS potential $\Vkseffs(\r)$ is composed of external, Hartree, and
exchange-correlation contributions,
\begin{equation}
 \Vkseffs(\r) = \Vext(\r) + \Vh(\r) + \Vxcs(\r),
\label{eq:V_KS}
\end{equation}
among which $\Vh(\r)$ and $\Vxcs(\r)$ depend on the charge
density (and wave functions for orbital-dependent $\Vxcs(\r)$
\cite{YHK}), hence Eqs.  (\ref{eq:KS-eq}),(\ref{eq:density}), and
(\ref{eq:V_KS}) form a self-consistent system of equations.  In the
usual local density approximation, $\Vxcs(\r)$ is calculated
inexpensively, hence the solution of KS equations
[Eq. (\ref{eq:KS-eq})] and generation of $\Vh(\r)$ comprise main parts
of calculations.  For localized systems, $\Vh(\r)$ are typically
obtained by solving the Poisson equation
\begin{equation}
 \nabla^2 \Vh(\r) = -4 \pi n(\r).
\label{eq:Poisson_eq}
\end{equation}
where $n(\r)$ is the total charge density
$n(\r)=n_{\downarrow}(\r)+n_{\uparrow}(\r)$.

In our higher-order finite-difference real-space formulation
\cite{Che94,IHL,YHK}, we discretize both KS and Poisson equations on a
three-dimensional uniform grid (with grid spacing $h$) with a
higher-order finite difference representation (with finite-difference
order $N$)
\begin{equation}
 \frac{{\mathrm{d}}^2}{{\mathrm{d}} x^2} f(x)
 = \sum_{j=-N}^{N} C_j f(x+jh) + O(h^{2N+2}),
\label{eq:FD}
\end{equation}
where $\{C_j\}$ are constants calculated by the algorithm of
Ref. \cite{For94}. KS equations have been solved by the preconditioned
conjugate gradient (CG) method \cite{Tet89,Byl90} supplemented by
subspace diagonalizations with {\it localized}\footnote{Vanishing
boundary condition is used for wavefunctions.  See
Sec. \ref{subsec:F90-OOP}.} wavefunctions. The Hartree potential has
been obtained from the solution of Poisson equation
[Eq. (\ref{eq:Poisson_eq})] on the {\it entire} simulation box, by
first generating boundary values with multipole expansion, and then
propagating solutions inside of the box with the combination of the
low-order finite-difference (N=1) fast Fourier transform method and
the higher-order finite-difference (typically N=5) preconditioned CG
method.  At each self-consistent step we generate a new input
Hartree-exchange-correlation potential using the simple linear mixing
of output and input potentials.  The computational procedure is
summarized in Fig. \ref{fig:flowchart}.
%
%
Further details of the computation method can be found in
Ref. \cite{IHL,YHK}.

\section{Fortran 90 implementation of object-oriented concepts}
\label{sec:F90-OOP}

\subsection{Object-oriented programming}
\label{subsec:OOP}

Roughly speaking, an object consists of a set of operations on some
hidden data.  Following Rumbaugh {\it et al.} \cite{Rum91}, the key
components of the \oo\ approach are:
\begin{enumerate}
\item {\em Identity} $-$
	data is quantized into discrete, distinguishable objects.
\item {\em Classification} $-$
	objects with the same data structure and behavior are grouped
	into a class.
\item {\em Polymorphism} $-$
	the same operation may behave differently on different
	classes.
\item {\em Inheritance} $-$
	sharing of data structures and behaviors among classes based
	on a hierarchical relationship.
\end{enumerate}
In addition, there are several themes underlying \oo\ technology which give
corresponding benefits:
\begin{itemize}
\item {\em Abstraction} $-$
	focusing on the essential, inherent aspects of an entity
	enhances the understanding of the problem itself, and
	preserves the freedom to make decisions as long as possible by
	avoiding premature commitments to details.
\item {\em Encapsulation} $-$
	separating the external aspects of an object from the internal
	implementation details of the object prevents a program from
	becoming so interdependent that a small change has massive
	ripple effects.
\item {\em Combining data and behavior} $-$
	keeping data structure hierarchy identical to the operation
	inheritance hierarchy shifts the burden of deciding what
	implementation to use from the calling code to the class
	hierarchy.
\item {\em Sharing} $-$
	sharing of code using inheritance induces savings in code and
	more importantly conceptual simplicity by reducing the number
	of distinct cases that must be understood and analyzed.
\item {\em Emphasis on object structure, not procedure structure} $-$
	stressing what an object {\it is}, rather than how is is {\it
	used}, makes the program more stable in the long run, since
	the features supplied by an object are much more stable than
	the way it is used as requirements evolve with time.
\item {\em Synergy} $-$
	using identity, classification, polymorphism, and inheritance
	together results in usually cleaner, more general, and more
	robust program.
\end{itemize}
These abstract ideas will be made concrete by examples in the next
section.

\subsection{Fortran 90 implementation of object-orientation}
\label{subsec:F90-OOP}

In this section we describe the \oo\ construction of grid-related
objects in \F\ 90.
\F\ 90 keywords will be denoted as bold uppercase characters, and
\oo\ concepts relevant to the discussion will be shown
in italic characters.  Before proceeding, we briefly consider the
modeling of objects: The most basic components in the
finite-difference electronic-structure code is the grid, which has the
information of grid coordinates and number of grid panels along the
$x-$,$ y-$, and $z-$directions.  We choose to use a uniform grid along
each direction.  In actual calculations, however, we employ only a
localized region in real space to save the memory and enhance the
computational efficiency.  In addition, the grid is apparently
constructed in a simulation box with basic information on the grid
starting and finishing coordinates.  So, we actually have a hierarchy
of three grid-related physical objects whose two-dimensional
representations as shown in Fig. \ref{fig:hierarchy}.
%
%
Below we present the implementations of this hierarchy of concepts
using \TYPE s of simulation box (\simbox), grid (\grid), and sub-grid
(\subgrid)\footnote{All the program listings in this paper have been
simplified from original versions for clarity of presentation.  Each
module includes more subroutines, and double-precision has been used
for real variables.}.

\begin{itemize}

\item[$\Box$]{\MODULE\ for \TYPE\ \simbox\ and corresponding procedures}
\begin{verbatim}

MODULE m_simbox
  IMPLICIT NONE
  PRIVATE
  PUBLIC :: simbox,new,display,...
  TYPE simbox
	 ! Initial/final coordinates of the simulation box
	 ! along x/y/z-dir.
	 REAL :: xi,xf,yi,yf,zi,zf
  END TYPE simbox
  INTERFACE new
	 MODULE PROCEDURE simbox_construct
  END INTERFACE
  INTERFACE display
	 MODULE PROCEDURE simbox_print
  END INTERFACE
  ...
CONTAINS
  SUBROUTINE simbox_construct(x1,x2,y1,y2,z1,z2,box)
  ! Assign given initial/final coordinates of the simulation
  ! box to 'simbox' components.
	REAL, INTENT(IN) :: x1,x2,y1,y2,z1,z2
	TYPE(simbox), INTENT(OUT) :: box
	...
  END SUBROUTINE simbox_construct
  SUBROUTINE simbox_print(box,name)
  ! Print out simulation box information.
	TYPE(simbox), INTENT(IN) :: box
	CHARACTER*(*), INTENT(IN), OPTIONAL :: name
	...
  END SUBROUTINE simbox_print
  ...
END MODULE m_simbox


\end{verbatim}

\item[$\Box$]{\MODULE\ for \TYPE\ \grid\ and corresponding procedures.}
\begin{verbatim}

MODULE m_grid
  USE m_simbox, ONLY: simbox
  IMPLICIT NONE
  PRIVATE
  PUBLIC :: grid,new,delete,display,...
  TYPE grid
	 ! Pointer to simulation box
	 TYPE(simbox), POINTER :: pt_simbox
	 ! Number of grid panels along x/y/z dir.
	 INTEGER  :: nx,ny,nz
	 ! Number of total grid points
	 INTEGER  :: ngrid
	 ! Grid spacings along x/y/z dir.
	 REAL :: dx,dy,dz
	 ! grid coordinates along x/y/z dir
	 REAL, DIMENSION(:), POINTER :: xcrd,ycrd,zcrd
  END TYPE grid
  INTERFACE new
	 MODULE PROCEDURE grid_construct
  END INTERFACE
  INTERFACE delete
	 MODULE PROCEDURE grid_destruct
  END INTERFACE
  INTERFACE display
	 MODULE PROCEDURE grid_print
  END INTERFACE
  ...
CONTAINS
  SUBROUTINE grid_construct(box,n1,n2,n3,grd)
  ! For the given simulation box, and the number of grid
  ! panels along each direction, assign/construct grid
  ! components.
	TYPE(simbox), INTENT(IN), TARGET :: box
	INTEGER,  INTENT(IN) :: n1,n2,n3
	TYPE(grid), INTENT(OUT) :: grd
	...
	! Simulation box coordinates
	grd%pt_simbox => box
	...
	! Grid coordinates generation.
	ALLOCATE(grd%xcrd(0:n1), ... STAT=ierr)
	IF(ierr/=0) ... ! Allocation error handling
	...
  END SUBROUTINE grid_construct
  SUBROUTINE grid_destruct(grd)
	TYPE(grid) :: grd
	...
	! Nullify pointer to 'simbox'
	IF(ASSOCIATED(grd%pt_simbox)) NULLIFY(grd%pt_simbox)
	! Deallocate 'xcrd','ycrd','zcrd'
	IF(ASSOCIATED(grd%xcrd)) THEN
	   DEALLOCATE(grd%xcrd, STAT=ierr)
	   IF(ierr/=0) ... ! Deallocation error handling
	ENDIF
	...
  END SUBROUTINE grid_destruct
  SUBROUTINE grid_print(grd,name)
  ! Print out grid information.
	USE m_simbox, ONLY: display
	TYPE(grid), INTENT(IN) :: grd
	CHARACTER*(*), INTENT(IN), OPTIONAL :: name
	...
	CALL display(box)
	...
  END SUBROUTINE grid_print
  ...
END MODULE m_grid


\end{verbatim}

\item[$\Box$]{\MODULE\ for \TYPE\ \subgrid\ and corresponding procedures.}
\begin{verbatim}

MODULE m_subgrid
  USE m_grid,   ONLY: grid
  IMPLICIT NONE
  PRIVATE
  PUBLIC :: subgrid,new,delete,...
  TYPE subgrid
	 ! Pointer to 'grid'
	 TYPE(grid), POINTER :: pt_grid
	 ! Number of sub-grid points
	 INTEGER :: nsubgrid
	 ! local grid index number,
	 ! 0 for outside of localized region.
	 INTEGER, DIMENSION(:,:,:), POINTER :: index
  END TYPE subgrid
  INTERFACE new
	 MODULE PROCEDURE subgrid_construct
  END INTERFACE
  INTERFACE delete
	 MODULE PROCEDURE subgrid_destruct
  END INTERFACE
  ...
CONTAINS
  SUBROUTINE subgrid_construct(grd,sbgrd)
	TYPE(grid), INTENT(IN), TARGET :: grd
	TYPE(subgrid), INTENT(OUT) :: sbgrd
	...
	! Assign pointer to full grid.
	sbgrd%pt_grid => grd
	! Allocate subgrid index array and initialize
	ALLOCATE(sbgrd%index(0:grd%nx,0:grd%ny,0:grd%nz), &
			 STAT=ierr)
	IF(ierr/=0) ! Deallocation error handling
	...
  END SUBROUTINE subgrid_construct
  SUBROUTINE subgrid_destruct(sbgrd)
	TYPE(subgrid) :: sbgrd
	...
	! Nullify pointer to 'grid', 'pt_grid'
	IF(ASSOCIATED(sbgrd%pt_grid)) NULLIFY(sbgrd%pt_grid)
	! Deallocate 'index' array of type 'subgrid' variable
	IF(ASSOCIATED(sbgrd%index)) THEN
	   DEALLOCATE(sbgrd%index,STAT=ierr)
	   IF(ierr/=0) ! Deallocation error handling
	ENDIF
  END SUBROUTINE subgrid_destruct
  ...
END MODULE m_subgrid


\end{verbatim}
\end{itemize}

First, note that in each case we define a new \TYPE\ which consists of
corresponding variables, such as {\bf xi,xf, etc.} for {\simbox} and
{\bf nx,ny,nz, etc.}  for {\grid}.  The ability to define derived
\TYPE s is a crucial ingredient of \oo\ code construction.  [{\em
abstraction}] By using these newly defined \TYPE s, we
construct/destroy corresponding variables together, and especially
pass them to procedures (\FUNCTION s and \SUBROUTINE s) as a single
argument, which enables the procedure interfaces to be simple and
stable.  [{\em identity}] We locate a \TYPE\ definition in the
corresponding \MODULE\ to make it globally accessible.  In addition,
note that we hide the implementation details by first declaring all
the entities in the
\MODULE\ as \PRIVATE\ and list only exceptions as
\PUBLIC\ for {\em data hiding}\footnote{ One can list all the
entities in each \MODULE\ as \PRIVATE\ or \PUBLIC, or make the default
\PUBLIC\ and then only list exceptions as \PRIVATE.  However, for the
purpose of {\em data hiding}, the current form is strongly
recommended \cite{Ell94}.}. [{\em encapsulation}] New \TYPE\ definitions
made \PUBLIC\ to outside can be \USE d in the calling routines by
including the corresponding module, and individual components of the
\TYPE\ can be accessed by following the variable by a percentage sign
$\%$ and the name of the component\footnote{ One can even hide the
data components of a derived \TYPE\ to the outside of the \MODULE\ by
preceding the first component declaration in the derived \TYPE\
definition by \PRIVATE\ attribute.  In that case, to access the
components, it is required to write procedures which manipulate and
return the components and include them in the same \MODULE.}.
Note that, for enhanced safety, we use \ONLY\ qualifier to access
public entites in the \USE d module.
%

Secondly, in all the \MODULE s, we \CONTAIN\ procedures which operate
on the corresponding \TYPE\ definition.  Hence when employing each
\MODULE, the user will access the data structure and its behavior at
the same time. [{\it classification}] Hence at this stage we have
achieved the basic requirements for the ``object-orientation'': we
organized program as collection of discrete objects that incorporate
both data structure (\TYPE) and behavior (procedures attached to the
\TYPE\ by being \CONTAIN ed).

Next, note that, when we \CONTAIN\ procedures, we employ
\MODULEPROCEDURE\ statement to give them generic names, and make those
generic names (instead of original names of the procedures) be
accessible from the outside by giving them a \PUBLIC\ attribute.  In
doing so, we can give different procedures a single generic name.  For
example, we use the same generic name \new\ for different procedures
of \TYPE s, \simbox\ (\simboxconstruct), \grid\ (\gridconstruct), and
\subgrid\ (\subgridconstruct).[{\em polymorphism\/}\footnote{ To be
more specific, this is {\it static polymorphism}.  A good discussion
on how to implement a {\it run-time polymorphism} can be found in
Ref. \cite{Dec98}.}]

Finally, in the \MODULE\ \Mgrid, \TYPE\ \grid\ inherits the
\simbox\ information, and again this \grid\ information
(including that of \simbox) is inherited to \TYPE\ \subgrid\ in
\MODULE\ \Msubgrid. [{\em inheritance}] So, \TYPE\ \grid\
variables will contain information on the simulation box ({\bf
xi,xf, etc}), and \TYPE\ \subgrid\ variables will contain the
information on the \grid\ ({\bf nx,ny,nz, etc.}) and the \simbox\
in which the \grid\ has been constructed.  Note that for the
implementation of this hierarchy structure, we have used \POINTER
s. In \F\ 90, to avoid execution efficiency degradation, all
objects to which a \POINTER\ may point should have a \TARGET\
attribute.  Hence, input variables (\INTENT\ attribute \IN) {\bf
box} in the procedure \gridconstruct\ (generic name \new) of the
\MODULE\ \Mgrid\ and {\bf grd} in the procedure
\subgridconstruct\ (generic name \new) of the \MODULE\ \Msubgrid\
have the \TARGET\ attribute.  \POINTER s have been associated
with \TARGET s by \POINTER\ assignment statements:

\hspace{2cm}    {\it pointer} $=>$ {\it target}

in the procedures \gridconstruct\ and \subgridconstruct.
In addition inheriting data, it is also possible for a procedure to
inherit another procedure.  In our example, a procedure \gridprint\ in
the \MODULE\ \Mgrid\ inherits (by \USE\ statement) another procedure
\simboxprint\ (generic name \display) located in the \MODULE\ \Msimbox.
Again, note that we give a generic name \display\ to both {\bf
grid$\_$print} and {\bf simbox$\_$print}.

Now we comment on additional \F\ 90 language features related with our
examples.  First, we do not use a \MODULE\ as a storage place of
global variables, although it is possible to do so, because we found
it is rather clumsy and risky for the large-scale coding due to the
problem of global storage similar to that arises in the usage of
COMMON block in \F\ 77.  \MODULE\ is exclusively used as a place for
\TYPE\ definitions and corresponding procedures.  Next, since
\ALLOCATABLE\ arrays cannot be used in derived \TYPE\ definitions, we
use instead \POINTER s (see \gridconstruct\ and \subgridconstruct).
Again, in this case of \POINTER\ usage, we arrange the allocation to
occur in a well-defined corresponding procedure contained in the same
module, which makes the usage of the derived \TYPE\ more safe and
robust.  Note that pointer disassociation (\NULLIFY) and deallocation
(\DEALLOCATE) are also handled in a similar way (see \griddestruct\
and pwd\subgriddestruct).

Before closing this section, we summarize the strategy of the modeling
of a new (or the remodeling of a present) large-scale code in \F\ 90
with \oo\ concepts, which we found useful.
\begin{enumerate}
\item Identify an object and define the corresponding variables as a
	  \TYPE.  These variables are typically global variables, or
	  frequently passed variables from the main program to procedures,
	  used together in a procedure in the conventional non-\oo\
	  programming style.  Be careful on the hierarchy (dependence) of
	  the objects.
\item Construct (identify) subroutines closely related with the \TYPE.
	  (Re)arrange the procedure interfaces (and contents if
	  necessary) using the \TYPE\ definition.
\item Define a \MODULE\ corresponding to the \TYPE, and include the
	  \TYPE\ definition from step 1 and \CONTAIN\ procedures identified
	  in step 2 in the \MODULE, and give them generic names.
\item Make the \TYPE\ definition and generic names of procedures \PUBLIC.
\end{enumerate}
Remind that the initial stage takes most of the time in the \oo\
approach, which is especially true in the remodeling of an existing
code, since the remaining process is mostly changing interfaces and
variable names in existing procedures and including them in \MODULE s
(assuming that the previous code is well-designed, hence the
restructuring is straightforward).  It should be also emphasized that
we find this process is actually beneficial for doing physics itself,
in that the programmer (usually the physicist herself or himself) has
to consider (reconsider in the case of remodeling of a present code)
and identify carefully the structures embedded in the problem under
consideration, hence results in making one focus more on the physical
pictures.  We refer the reader to Ref. \cite{Rum91} for further
discussion of the benefits of \oo\ programming.

\section{Object-oriented implementation of multigrid methods}
\label{sec:OOP-MG}

\subsection{Multigrid method}
\label{subsec:MG}

We first briefly review the theory of the \mg\ method
\cite{Hea97,Pre92,Bri87,Dou96}. The fundamental idea behind all
\mg\ methods is to combine computations done on different scales,
based on the observation that many iterative methods tend to reduce
the high-frequency (i.e. oscillatory) components of the error rapidly
but reduce the low-frequency (i.e. smooth) components of the error
much slowly. Since the notions of smooth or oscillatory components of
the error are relative to the mesh on which the solution is defined,
in particular, a component that appears smooth on a fine grid may
appear oscillatory on a coarse grid, we can naturally think of using
coarser grid to reduce this (now oscillatory) error and interpolate
back to the fine grid. The key to the successful implementation of a
\mg\ method is the choice of grids of different scales, the strategy
to proceed through them, and how we move objects among them. Broadly
the \mg\ algorithms can be divided into two basic categories
\cite{Dou96}:
\begin{enumerate}
\item Correction methods $-$ start at the finest level, and use
		the coarser levels solely to compute a correction which is
		added to the approximate solution on the finest level.
\item Nested iteration methods $-$ generate initial guesses on coarser levels
		and frequently reuse coarser levels for corrections also.
\end{enumerate}
Specific examples of these correction path and nested iteration path
are shown in Fig \ref{fig:MG}-(a) and (b).
%
%
Roughly speaking, when a good initial guess is available, a correction
algorithm can be used, otherwise we should use a nested iteration
scheme \cite{Dou96}.

\subsection{One-way multigrid method in electronic-structure calculations
and its implementation}
\label{subsec:OW-MG}

Multigrid is a quite general concept, and apparently the choice of a
specific algorithm depends on the nature of the problem under
consideration.  We have recently demonstrated that the introduction of
a simple one-way \mg\ method (Fig. \ref{fig:MG}-b) greatly improves
the efficiency of real-space electronic-structure calculations based
on the iterative solution of KS equations \cite{IHL3}. The motivation
of our work was based on the observation that the most time-consuming
part of the self-consistent electronic-structure calculations
described in Sec. \ref{sec:KS-DFT} is the iterative solution of KS
equations \cite{Tet89,Byl90}.  The sources of this computation
bottleneck can be traced to broadly two (but closely related) aspects
of self-consistent iterative diagonalization schemes.  First of all,
in general we do not have a good initial guess of wave functions,
which generate density, and hence $\Vh(\r)$ and $\Vxcs(\r)$ in
Eq. ({\ref{eq:KS-eq}).  So initial several self-consistency steps will
be used to obtain solutions of biased Hamiltonians, although they tend
to be the most time-consuming part.  Secondly, in single iterative
solution of KS equations, a direct application of a relaxation method
on the fine grid has trouble in damping out the long-ranged or slowly
varying error components in the orbitals. This can be understood by
the usual spectral analysis of relaxation scheme\cite{Bri87}, or
considering that the nonlocal Laplacian operation on a fine grid is
physically short-ranged.

Hence, for our purpose, we seek a \mg\ procedure which generates a
good initial guess for the finest grid calculation and effectively
removes long-range error components of wave functions in the solution
of KS equations [Eq. (\ref{eq:KS-eq})].  While an efficient
interpolation/projection scheme is a crucial ingredient of any
successful application of \mg\ method, we note that it can be also
time-consuming and tricky part since in our case we need to transfer a
large number of wave functions which should observe the orthonormality
conditions.  Hence our strategy, which is the characteristic of the
scheme, is to minimize the number of data transfer between different
grid levels, and employ an accurate interpolation method which is very
accurate and allow us to use even a noninteger ratio of grid spacings:
The calculation starts from the coarsest grid $2h$, and in each
grid-level calculation, Eqs. (\ref{eq:KS-eq}) and
(\ref{eq:Poisson_eq}) are solved self-consistently as in the usual
single-level algorithm shown in Fig. \ref{fig:flowchart}.  After each
self-consistent calculation on a coarse grid, only wave functions are
interpolated to the next fine grid, and another set of self-consistent
calculation is performed.  Since that the interpolated wave functions
usually do not satisfy the orthonormality condition any more, we take
an extra Gram-Schmidt orthogonalization process after each orbital
interpolation.  Hence we have $n-1$ interpolations and Gram-Schmidt
orthogonalization processes for the $n-$level multigrid calculations.
For the interpolation, we used a three-dimensional piecewise
polynomial interpolation with a tensor product of one-dimensional
B-splines as the interpolating function\cite{Pre92,deB84}. A piecewise
cubic polynomials have been taken as B-splines. In
Fig. \ref{fig:OW-MG}, we summarize the computational procedure for the
case of three grids, $2h$, $1.5h$, and $h$. We refer the reader to
Ref. \cite{IHL3} for further discussion of the method and the
comparison with other \mg\ schemes \cite{Bri95,Anc99,Wan99}.
%

Now we describe the implementation of the one-way \mg\ algorithm of
Fig. \ref{fig:OW-MG} using the \oo\ components constructed in
Sec. \ref{subsec:F90-OOP}.  The main part of the code is shown below.

\begin{verbatim}
  SUBROUTINE mg_ks_dft(box,ng,...)
	USE m_simbox, ONLY: simbox
	USE m_grid, ONLY: grid,new,display,delete
	USE m_subgrid, ONLY: subgrid,new,delete
	USE m_wavefunction, ONLY: wavefunction,new,delete
	...
	IMPLICIT NONE
	TYPE(simbox), INTENT(IN) :: box ! Simulation box information
	INTEGER, INTENT(IN) :: ng(3) ! Number of grid panels
								 ! at the finest grid
	...
	! Local variables
	TYPE(grid), ALLOCATABLE, DIMENSION(:) :: grd
	TYPE(subgrid), ALLOCATABLE, DIMENSION(:) :: sbgrd
	TYPE(wavefunction), ALLOCATABLE, DIMENSION(:), TARGET :: wf
	...

	! Allocatable arrays
	! 'nlevel' is number of multigrid level
	ALLOCATE(grd(nlevel),sbgrd(nlevel),wf(nlevel),STAT=ierr)
	IF(ierr/=0) ...  ! Error handling
	...
	! One-way multigrid loop
	DO ilvl = 1,nlevel

	   ! Assign number of grid panels for each multigrid level
	   ! 'ratio' is grid spacing ratio with finest grid as 1
	   n1 = NINT(ng(1)/ratio(ilvl))
	   n2 = NINT(ng(2)/ratio(ilvl))
	   n3 = NINT(ng(3)/ratio(ilvl))

	   ! Generate grid for level='ilvl'
	   CALL new(box,n1,n2,n3,grd(ilvl))
	   CALL display(grd(ilvl))

	   ! Generate subgrid for level='ilvl'
	   CALl new(grd(ilvl),sbgrd(ilvl))

	   ! Effective region allocation in the subgrid
	   ...

	   ! Generate wavefunctions
	   CALL new(...,wf(ilvl))

	   ! If level>1, interpolate w.f.(ilvl-1) => w.f.(ilvl)
	   IF(ilvl>=2) THEN

		  ! Interpolate: wf(ilvl-1) => wf(ilvl)
		  ...
		  ! and orthonormalize them
		  ...

		  ! Destruct: grid(ilvl-1), subgrid(ilvl-1), wf(ilvl-1).
		  CALL delete(grd(ilvl-1))
		  CALL delete(sbgrd(ilvl-1))
		  CALL delete(wf(ilvl-1))
	   ENDIF

	   ! Adjust calculation parameters according to grid level.
	   ...

	   ! Solve KS-equations for level='ilvl'
	   CALL ksdft(...)

	   ! Remove grid(ilvl), subgrid(ilvl), wf(ilvl)
	   ! at the last MG step
	   IF(ilvl == nlevel) THEN
		  CALL delete(grd(ilvl))
		  CALL delete(sbgrd(ilvl))
		  CALL delete(wf(ilvl))
	   ENDIF

	END DO

	! Deallocate arrays
	DEALLOCATE(grd,sbgrd,wf)
	...

  END SUBROUTINE mg_ks_dft


\end{verbatim}

The simplification of the coding induced by \oo\ programming style
should be obvious in this example.  Since the simulation box is
assigned only once in the current method, it has been generated once
in the higher level and passed as an input variable, and only grids
and sub-grids built in the simulation box have been \ALLOCATE d for
the number of \mg\ levels\footnote{However, \simbox\ objects can be
also frequently generated if we perform adaptive mesh refinement type
calculations.}.  In addition to the objects we described in
Sec. \ref{subsec:F90-OOP}, we use \TYPE\ \wavefunction\ which has not
been described but has been constructed in a similar way as others.
Again, note that we use the same generic names of procedures for
different \TYPE s, \new\ and \delete.  It should be also noted that
actual allocatable arrays in \TYPE\ \grid\ ({\bf xcrd, etc.}),
\subgrid\ ({\bf index}), and most importantly \wavefunction\ whose
size is (number of grid points) $\times$ (number of states) $\times$
(number of spins) are only \ALLOCATE d when they are required, and
those arrays are \DEALLOCATE d as soon as they become unnecessary.
These processes are elegantly handled by \new\ and \delete\ calls.

\subsection{Performance test}
\label{subsec:performance}

For the analysis of performance enhancement due to our \mg\ method, we
reconsider a quasi-two-dimensional quantum dot model that has been
employed in Ref. \cite{IHL3}, in which a 20-electron quantum dot has
been studied with one-level ($h$), two-level ($2h$ and $h$), and
three-level ($2h$, $1.5h$, and $h$) methods.  Here, a more detailed
analysis of performance is provided, with varying number of electrons
up to 24 and additional four-level ($4h$, $2h$, $1.5h$, and $h$)
calculations.  While calculations were performed on the entire
simulation box with the original code in Ref. \cite{IHL3} , here we
employ the newly designed \oo\ code which use only the grid points
inside of a spherical region with a radius 8.0 $a_B^*$.  Two
calculations are further different in simulation parameters.

Quantum dot in GaAs host material (dielectric constant
$\epsilon=12.9$, effective mass $m^*= 0.067m_e$) is modeled by an
anisotropic parabolic confinement potential $\Vext(\r)= \frac{1}{2}
\omega_x^2 x^2 +\frac{1}{2} \omega_y^2 y^2 +\frac{1}{2} \omega_z^2 z^2$,
in which the $z$-axis is taken as the dot growth direction.  As in
Ref. \cite{IHL3}, we use the confinement energies
$\omega_x=\omega_y=5$ meV, and $\omega_z=45$ meV.  Our calculations
are based on the effective mass approximation, and rescaled length and
energy units are respectively $a_B^*$ = 101.88 $\AA$ and 10.96 meV.
Uniform grid spacing $h=0.3a_B^*$ with box size $18 \times 18 \times
18 a_B^{* \ 3}$ have been used.  Incorporation of the spherical local
region results in the usage of only about $35\%$ of total number of
grids, hence the number of grid points involved in the calculations is
$1.2 \times 10^3$ for grid $4h$, $1.0 \times 10^4$ for grid $2h$, $2.4
\times 10^4$ for grid $1.5h$, and $7.9 \times 10^5$ for grid $h$.
Finite-difference order $N$ [Eq. (\ref{eq:FD})] for the solution of
the KS equations and Poisson equation are chosen such that the range
of the physical coverage is approximately same, so $N=5$ for $h$,
$N=3$ for $1.5h$ and $2h$, and $N=1$ for $4h$.  Noninteracting
eigenstates (Hermite polynomials) are used as an initial guess for the
coarsest grid calculation.  Spin-unpolarized calculations have been
performed for the simplicity of performance comparison, although the
spin-polarized scheme should be employed to observe possible
spin-polarized states and the corresponding Hund's rule \cite{IHL}.

We first show the CPU times of self-consistent iterations in 1-, 2-,
3-, and 4-level 24-electron calculations in Fig. \ref{fig:CPU} to
contrast the characteristics of self-consistent calculations in the
conventional 1-level and \mg\ methods.
%
%
The horizontal axis stands for the self-consistency iteration index,
while the vertical axis is the required computer time for a given
iteration step.  Interpolation and orthonormalization steps in
multi-level calculations are indicated by downward arrows.  While the
\mg\ calculations requires more number of self-consistent iterations
in general, they are mostly performed in the coarsest grid, and at the
finest grid level $h$ we only need two or three iterations, which
demonstrates that coarse grid calculations provide a good initial
solution for the finest grid $h$ calculation and results in
significant time saving.

In Fig. \ref{fig:levels}, we compare the performance of different \mg\
strategies for different number of electrons.
%
%
Note that in general the use of \mg\ improves the computation speed,
and moreover its efficiency increases with the system
size. Computation speed-up defined as (CPU time for 1-level
calculation)/(CPU time for $n$-level calculation) amounts to more than
7 for the 24 electron case with 4-level method.  Second, while the 3-
and 4-level computations are usually better than the 2-level one, its
specific performance varies with the number of electrons.  The rule of
thumb is that 4-level method should be used for the electron number
larger than 20, otherwise 3-level is sufficient.

\section{Conclusions}
\label{sec:conclusions}

In the modern computation era when the increase of computational
capability increases almost exponentially with time, it is clear that
physicists can attack more ambitious problems requiring more
challenging large-scale computations. However, with the growth of the
size of the problem, typically the complexity of the problem itself,
hence the complication of the code also increases. \Oo\ methodology
can be a valuable solution to this problem of complexity of modern
scientific computations, and, in this paper, we showed one example of
the application of the object-orientation methodology to the
large-scale code implementation in \F\ 90.  Specifically, we treated a
real-space grid-based electronic-structure program which solves the KS
and Poisson equations self-consistently, and especially explained how
we have implemented the one-way \mg\ method we have recently proposed
\cite{IHL3} using the \oo\ techniques.
According to our experience, we believe that it pays to write a
scientific program in \oo\ fashion in the long run, and further the
cost we have to pay is minimal compared with its benefits even when
using a non-\oo\ language \F\ 90.

\acknowledgments
This work has been supported by the National Science Foundation under
Grant No. DMR 9802373 (Y.-H.K. and R.M.M).  Computations were
performed in the Material Research Laboratory Center for Computation.



\newpage

\begin{figure}
\caption{Flowchart of the current higher-order finite-difference
electronic-structure calculations based on the iterative CG
diagonalization. Only the dotted parts are repeated in the
higher-level \mg\ calculations shown in Fig. \ref{fig:OW-MG}.}
\label{fig:flowchart}
\end{figure}

\begin{figure}
\caption{Two-dimensional representation of the hierarchy of three physical
	objects for grid-based calculations: (a) simulation box, (b)
	grid, and (c) sub-grid. Only filled circles in (c) are
	actually used for computations.}
\label{fig:hierarchy}
\end{figure}

\begin{figure}
\caption{Examples of multigrid algorithmic flow: (a) correction path,
	and (b) nested iteration path.  Level 3 has the finest grid,
	level 1 the coarsest; computation flows from left to right.}
\label{fig:MG}
\end{figure}

\begin{figure}
\caption{ Schematic diagram of the present one-way \mg\ algorithm
	for the case of three-level ($2h$, $1.5h$, and $h$)
	calculations.  The calculation starts at the coarsest level
	(level 1, $2h$) at the bottom, and ends at the finest grid
	(level 3, $h$) at the top.  At level 2 and 3, only the dotted
	parts of the self-consistent calculation in
	Fig. \ref{fig:flowchart} are performed.  Orbital interpolation
	and orthogonalization step is taken after each coarse grid
	(level 1 and 2) calculation.}
\label{fig:OW-MG}
\end{figure}

\begin{figure}
\caption{CPU time vs. self-consistent iteration number of
	twenty-four-electron quantum dot calculations in (a) one-level
	($h$), (b) two-level ($2h$ and $h$), (c) three-level ($2h$,
	$1.5h$, and $h$), and (d) four-level ($4h$, $2h$, $1.5h$, and
	$h$) schemes.  Downward arrows in (b), (c), and (d) indicate
	interpolation-orthonormalization steps. Total computation time
	is (a) 59.5, (b) 12.9, (c) 11.0, and (d) 8.3 minutes.
	Calculations are performed on DEC alpha 500au workstations.}
\label{fig:levels}
\end{figure}

\begin{figure}
\caption{Comparison of the computational efficiency enhancement in
	$n-$level one-way \mg\ methods, where $n$ is 2 ($2h$ and $h$),
	3 ($2h$, $1.5h$, and $h$), and 4 ($4h$, $2h$, $1.5h$, and
	$h$), for electron number 8, 12, 16, 20, and 24.  Computation
	speed-up has been defined as (CPU time for 1-level
	calculation)/(CPU time for $n$-level calculation). Total
	computation time of 1-level calculation has been denoted in
	minutes for each electron number. Calculations are performed
	on DEC alpha 500au workstations.}
\label{fig:CPU}
\end{figure}

\newpage
\begin{center}
\begin{minipage}[H]{\linewidth}
\vspace{1cm}
	\includegraphics[height=\linewidth]{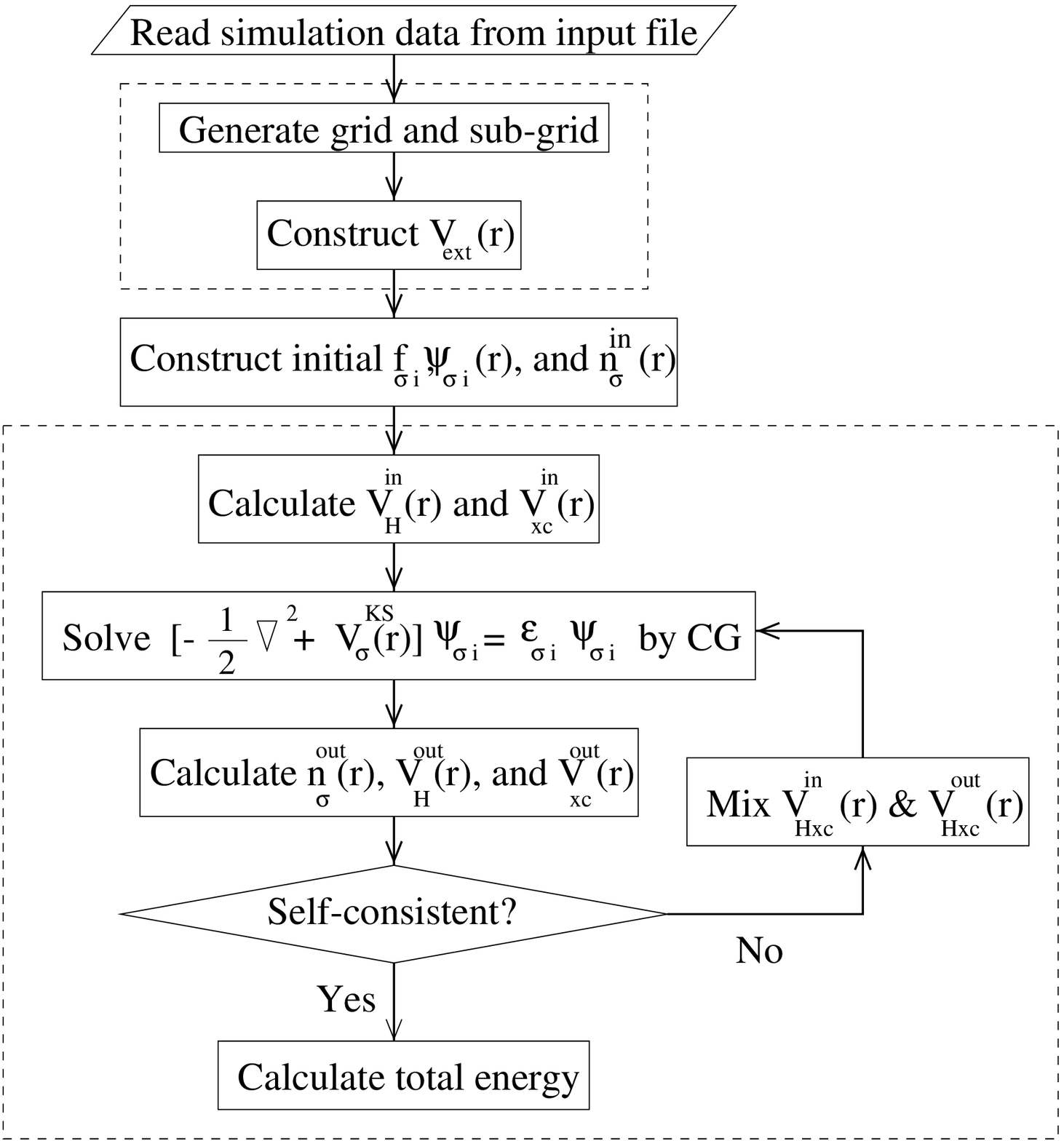}
\vspace{1cm}
\end{minipage}  \\ 
Fig. 1
\end{center}

\newpage
\begin{center}
\begin{minipage}[H]{\linewidth}
\vspace{1cm}
	\includegraphics[height=0.3\linewidth,angle=-90]{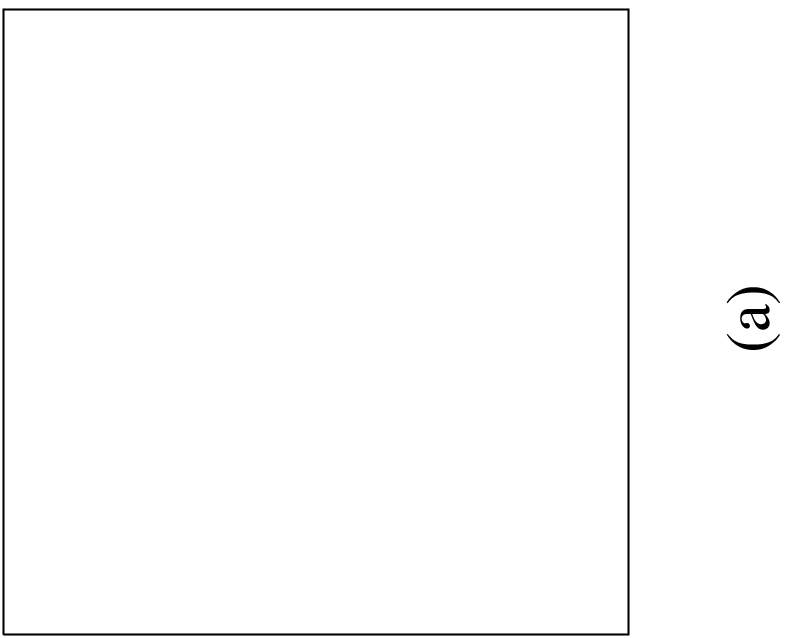}\hspace*{0.5cm}
	\includegraphics[height=0.3\linewidth,angle=-90]{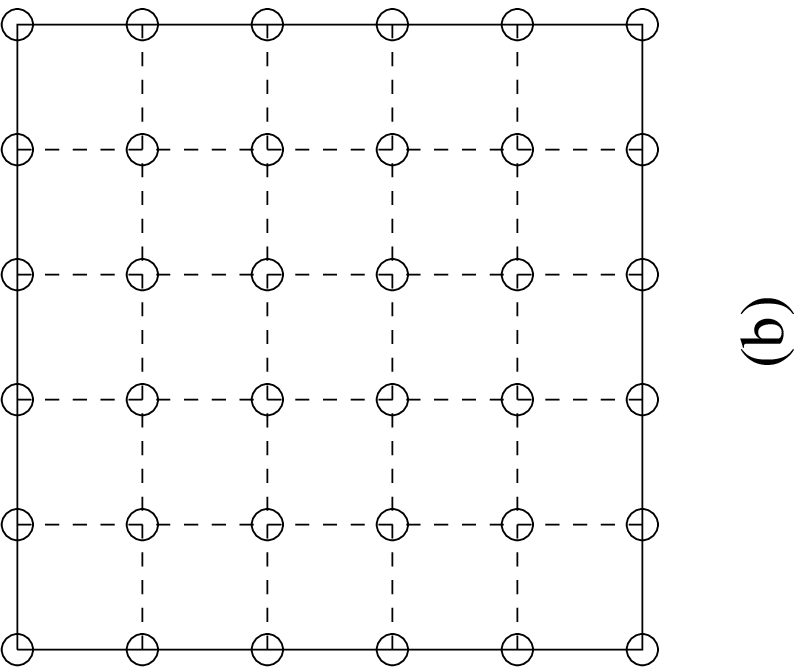}\hspace*{0.5cm}
	\includegraphics[height=0.3\linewidth,angle=-90]{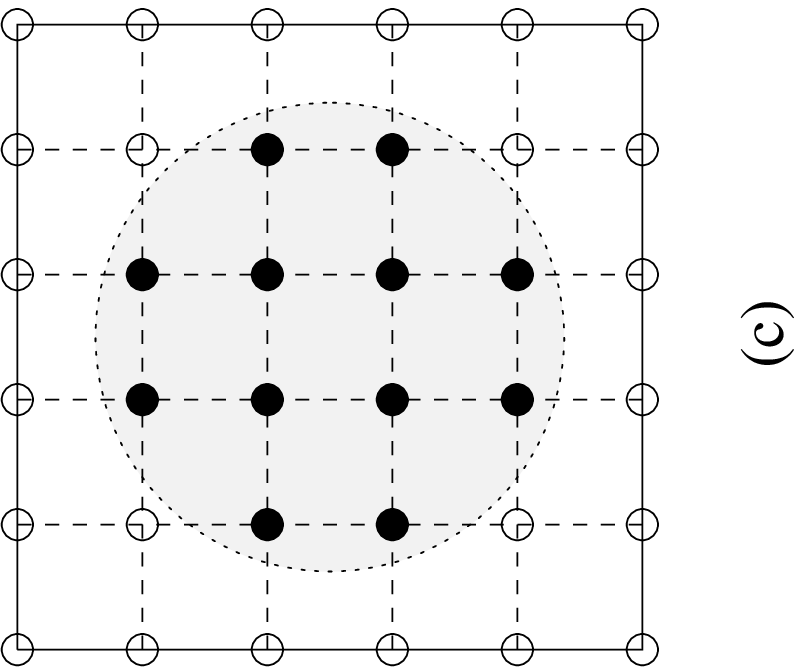}
\vspace{1cm}
\end{minipage}  \\
Fig. 2
\end{center}

\newpage
\begin{center}
\begin{minipage}[H]{1.00\linewidth}
\vspace{1cm}
	\includegraphics[height=\linewidth,angle=-90]{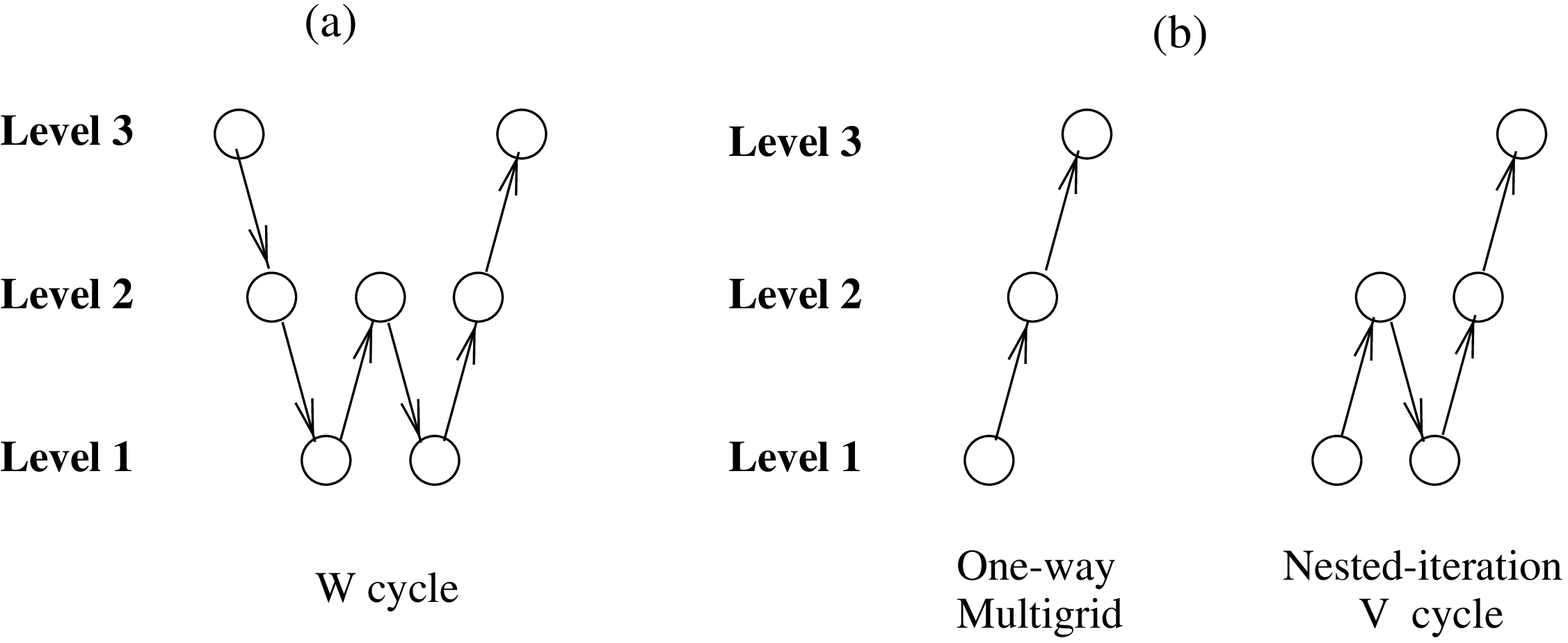}
\vspace{1cm}
\end{minipage}  \\
Fig. 3
\end{center}

\newpage
\begin{center}
\begin{minipage}[H]{1.00\linewidth}
\vspace{1cm}
	\includegraphics[height=\linewidth,angle=-90]{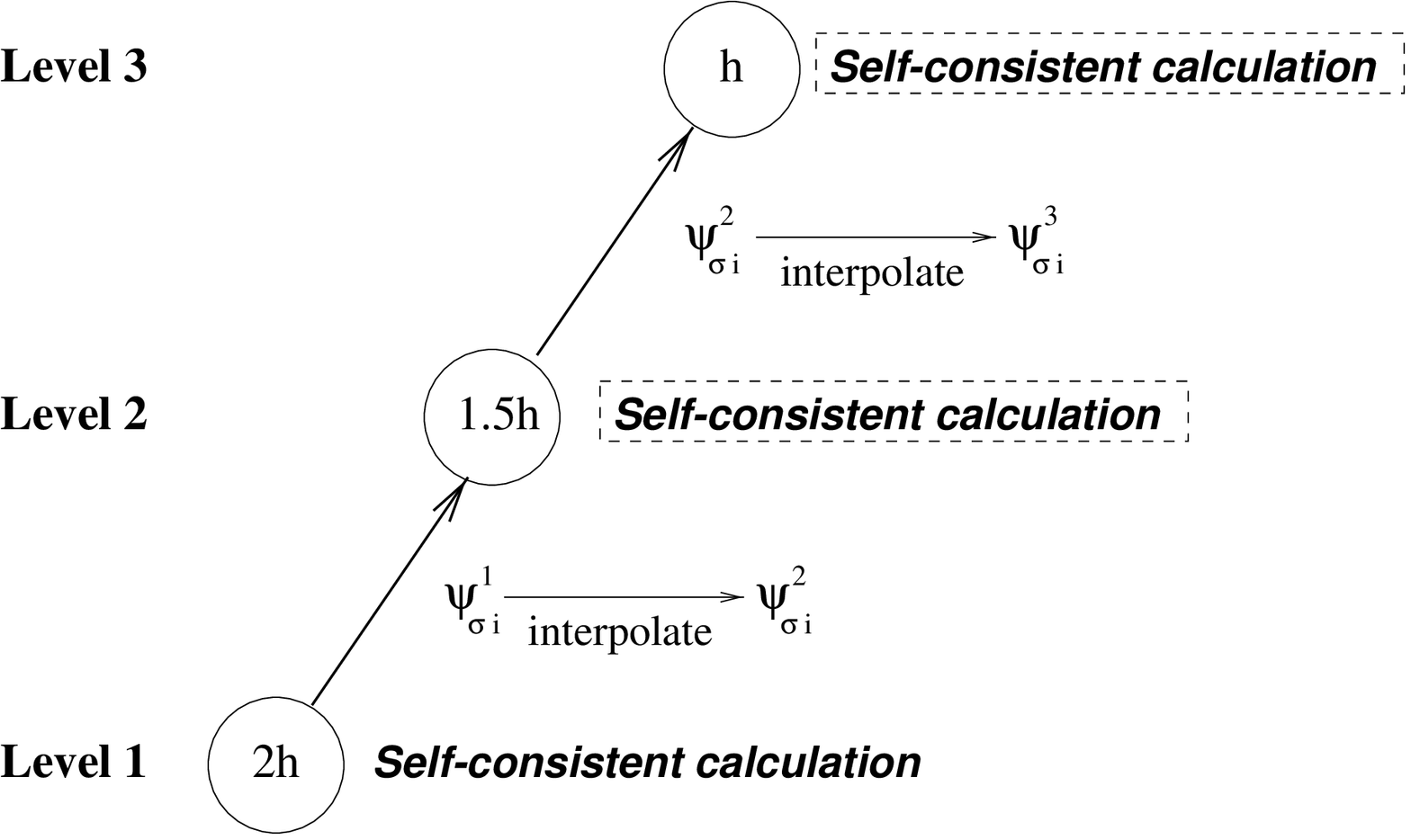}
\vspace{1cm}
\end{minipage}  \\
Fig. 4
\end{center}

\newpage
\begin{center}
\begin{minipage}[H]{1.00\linewidth}
\vspace{1cm}
	\includegraphics[height=\linewidth,angle=-90]{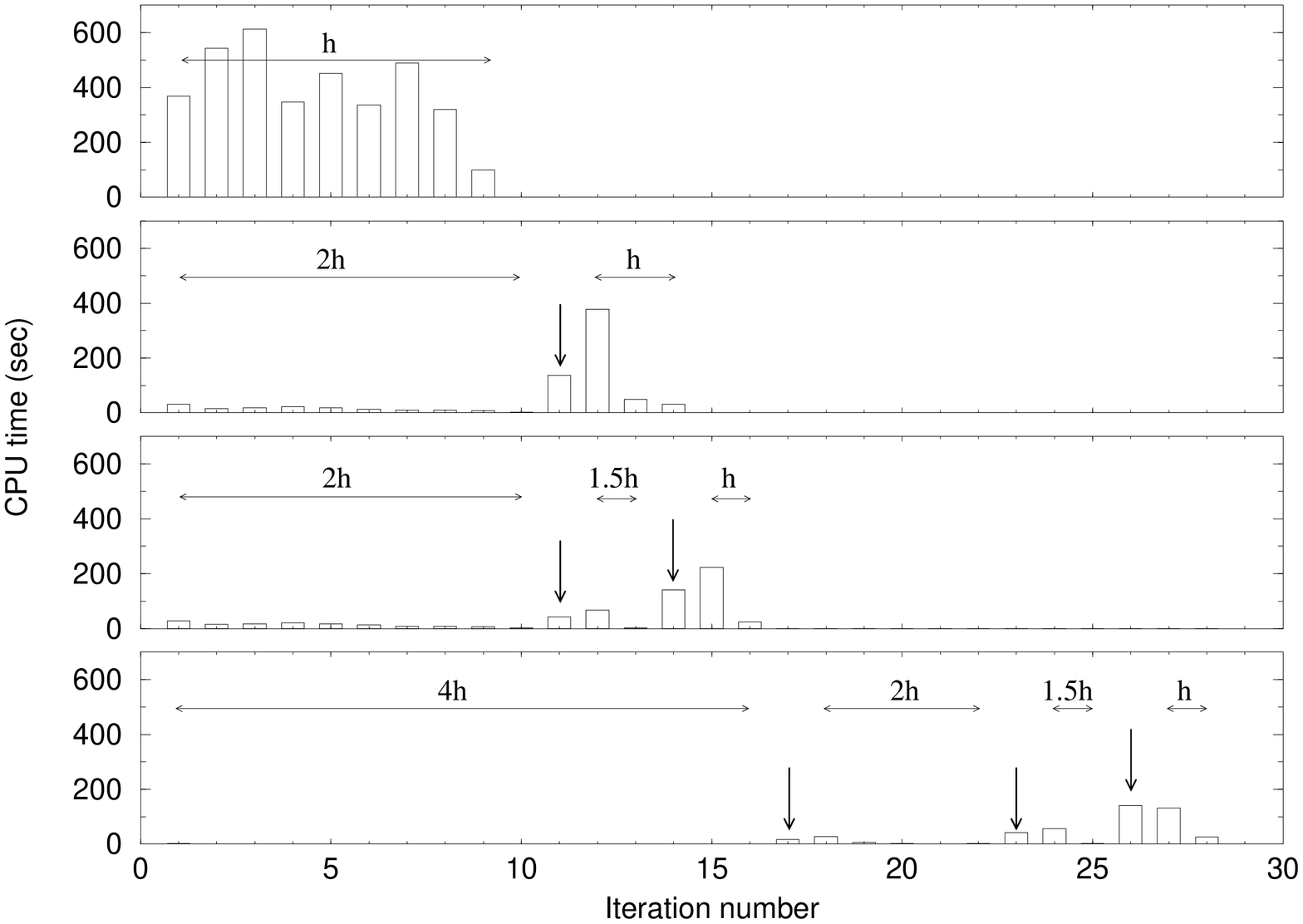}
\vspace{1cm}
\end{minipage} \\
Fig. 5
\end{center}

\newpage
\begin{center}
\begin{minipage}[H]{1.00\linewidth}
\vspace{1cm}
	\includegraphics[height=\linewidth,angle=-90]{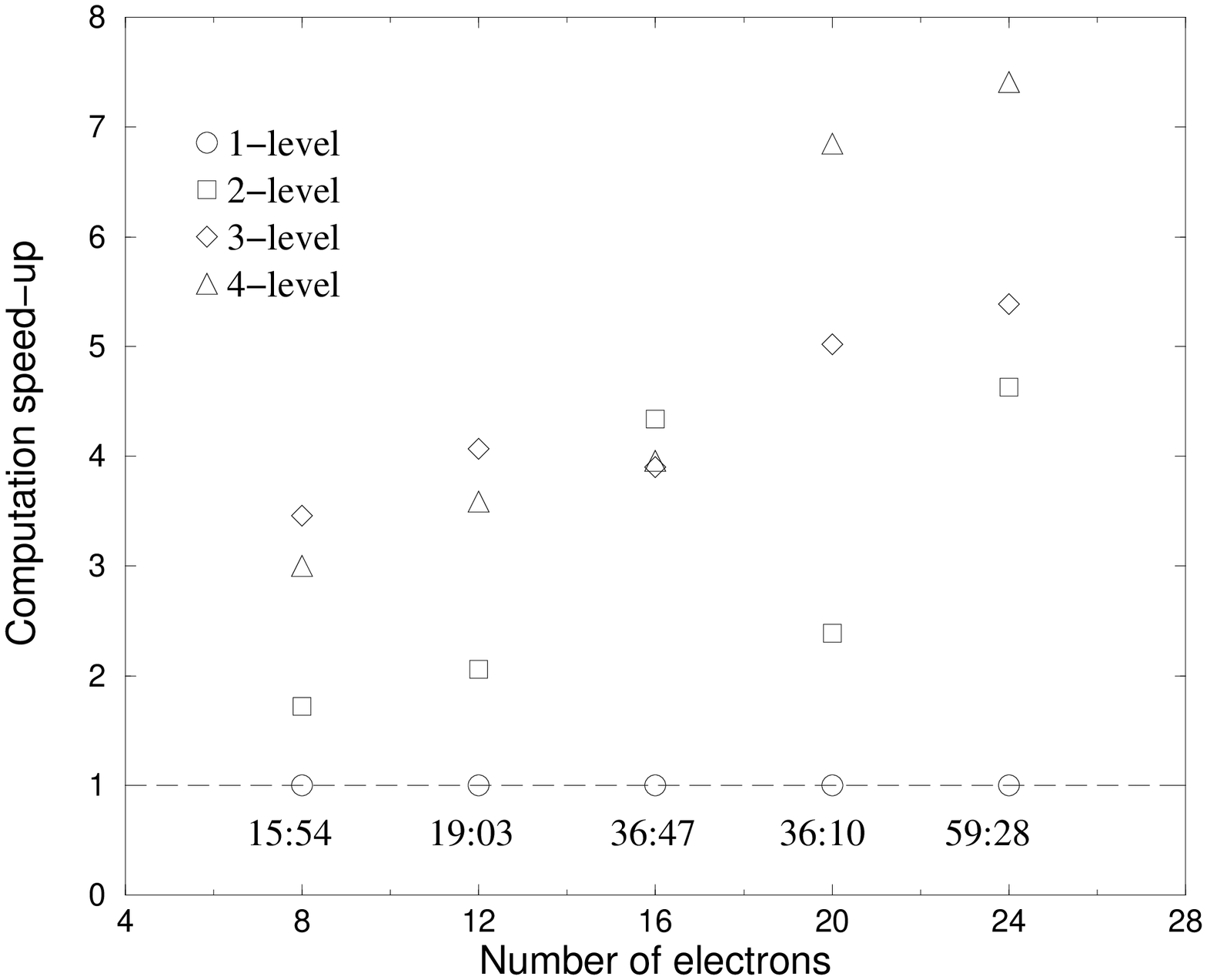}
\vspace{1cm}
\end{minipage} \\
Fig. 6
\end{center}

\end{document}